\documentclass[pra,twocolumn,amsmath,amssymb,superscriptaddress,eqsecnum,nofootinbib]{revtex4-2}
\usepackage{graphicx,amsmath,relsize,epstopdf,color,mathtools,bm,newtxtext,newtxmath,braket,rotating}
\usepackage[hyphenbreaks]{breakurl}
\usepackage[colorlinks=true,linkcolor=blue,citecolor=blue,urlcolor =blue]{hyperref}
\usepackage{soul}

\usepackage[table,xcdraw]{xcolor}

\begin{document}

\title{The quantum sky of Majorana stars}

\author{Luis~L.~S\'{a}nchez-Soto}
\affiliation{Departamento de \'{O}ptica, Facultad de F\'{\i}sica, Universidad Complutense, 28040~Madrid, Spain}
\affiliation{Max-Planck-Institut f\"{u}r die Physik des Lichts, 91058~Erlangen, Germany}
\affiliation{Institute for Quantum Studies, Chapman University, Orange, CA~92866, USA}

\author{Andrei B. Klimov}
\affiliation{Departamento de F\'{\i}sica, Universidad de Guadalajara, 44420~Guadalajara, Jalisco, Mexico}

\author{Aaron~Z.~Goldberg}
\affiliation{National Research Council of Canada, 100 Sussex Drive, Ottawa, Ontario K1N 5A2, Canada}

\author{Gerd Leuchs}
\affiliation{Max-Planck-Institut f\"{u}r die Physik des Lichts, 91058~Erlangen, Germany}
\affiliation{Institut f\"ur Optik, Information und Photonik, Universit\"at Erlangen-N\"{u}rnberg, 91058 Erlangen,  Germany}
\affiliation{Department of Physics, University of Ottawa, Ottawa, {Ontario K1N 6N5}, Canada}

\begin{abstract}
Majorana stars, the $2S$ spin coherent states that are orthogonal to a spin-$S$ state, offer an elegant method to visualize quantum states.  This representation offers deep insights into the structure, symmetries, and entanglement properties of quantum states, bridging abstract algebraic formulations with intuitive geometrical intuition. In this paper, we briefly survey the development and applications of the Majorana constellation, exploring its relevance in modern areas of quantum information.  
\end{abstract}
\maketitle

\setstcolor{red}

\section{Introduction}

In 1932, Ettore Majorana proposed a striking construction in which an arbitrary spin-$S$ state could be represented as a superposition of $2S$ spins $1/2$ (or qubits, in the language of modern quantum information)~\cite{Majorana:1932aa}. Written in Italian, his paper attracted little attention and might have been forgotten altogether, had it not been stumbled upon in 1935 by a 17-year-old Julian Schwinger. What caught Schwinger's attention was not Majorana's stated goal---an ambitious attempt to generalize the Stern–Gerlach experiment to magnetic fields and moments of arbitrary form---but a deceptively simple line in the text: ``...ogni stato sarà rappresentato da $2S$ punti sulla sfera unitaria." Majorana offered barely four lines of explanation, as if the statement were self-evident. Schwinger, intrigued, set about working it out for himself and soon confirmed its validity~\cite{Schwinger:1937aa}, though he chose not to dwell on its deeper implications~\cite{Schwinger:1977aa}. 

Majorana disappeared in 1938, at the age of 31.\footnote{Majorana vanished at sea in 1938 under circumstances that remain shrouded in mystery. His disappearance later gained wide public attention through Leonardo Sciascia’s  1975   novel \emph{The Vanishing of Majorana}, which elevated the episode to the level of  cultural myth and provoked a host of speculative interpretations regarding his motives.} In a parallel to his personal fate, his paper and its insights also remained obscure.  Schwinger returned to the subject in 1947, stimulated by a review  article  in which Bloch and Rabi~\cite{Bloch:1945aa} ``remarked on the derivation of Majorana’s formula from the spin 1/2 representation." Only in 1952~\cite{Schwinger:1952aa} did he provide a full treatment, introducing what is now called the Schwinger map, and showing how the commutation relations of angular momentum reduce elegantly to those of a two-dimensional isotropic harmonic oscillator. Surprisingly, this (unpublished) monograph cites works by Weyl, G\"{u}ttinger, van der Waerden, Racah and Wigner, but contains no mention of the Majorana formula, which was subsequently revisited by Meckler~\cite{Meckler:1958aa} yet passed over in Schwinger’s own account. 

In his paper, Majorana calculated the probability of nonadiabatic transitions occurring when an oriented atomic beam passes near a point where the magnetic field vanishes, a problem previously examined by G\"{u}ttinger~\cite{Guttinger:1932aa}. He made a remarkable prediction: even in the absence of a magnetic field, a spin-$1/2$ particle has a nonzero probability of flipping its spin. Bloch and Rabi later generalized this result to arbitrary spin-$S$ systems~\cite{Bloch:1945aa}, laying the theoretical groundwork for nuclear magnetic resonance.
Decades later, the same effect reemerged in atomic physics~\cite{Inguscio:2006aa}, where atoms confined in quadrupole magnetic traps were found to undergo spin flips and escape—a phenomenon known as the Majorana hole. This process, directly reflecting Majorana’s original prediction, initially hindered evaporative cooling and the realization of Bose–Einstein condensation. The issue was eventually resolved by developing trap configurations with a nonzero minimum magnetic field~\cite{Cornell:2002aa,Ketterle:2002aa}.

The Majorana representation gained renewed prominence through the work of Roger Penrose. In 1960, he introduced a spinor approach to general relativity~\cite{Penrose:1960aa}, providing a proof of Petrov’s classification of gravitational fields based on the degeneracy of the principal null directions of a gravitational spinor~\cite{Penrose:1984aa}. Decades later, recognizing these directions as Majorana constellations, he employed the representation in a refined proof of Bell’s theorem~\cite{Zimba:1993aa} and popularized it through his influential book~\cite{Penrose:1989aa}.

Subsequent research quickly expanded the scope of the representation. Hannay derived the pair correlation function of random spin states in the large-$S$ limit~\cite{Hannay:1996wl} and later obtained a general expression for Berry’s phase, applying the spin-1 case to optical polarization~\cite{Hannay:1998vb,Hannay:1998ab}. 

The geometric insight offered by the Majorana picture has since inspired numerous applications. In quantum information, it has been used to characterize ``the most quantum" states~\cite{Zimba:2006fk,Giraud:2010db,Giraud:2015oj,Bjork:2015ux,Bjork:2015ab,Goldberg:2020aa} and to develop powerful criteria of quantumness, such as the stellar rank~\cite{Chabaud:2020aa,Chabaud:2021aa,Motamedi:2026aa}.

The representation also provides a natural framework for studying entanglement. Bastin and collaborators classified symmetric $N$-qubit states under stochastic local operations and classical communication (SLOCC) using the degeneracy patterns of Majorana stars~\cite{Bastin:2009aa,Mathonet:2010aa}. Markham and coworkers~\cite{Aulbach:2010aa,Markham:2011aa} further related entanglement measures to constellation symmetries, while more recent work has  explored  multipartite entanglement from this perspective~\cite{Ribeiro:2011ti,Ganczarek:2012aa,Devi:2012aa,Mandilara:2014aa,Liu:2016aa,Kam:2020aa,Chryssomalakos:2021aa,Goldberg:2022ab,Rudzinski:2024aa}.

In quantum metrology, the representation has revealed the states most sensitive to small rotations around arbitrary axes~\cite{Bouchard:2017aa,Chryssomalakos:2017aa,Goldberg:2018aa,Martin:2020aa,Goldberg:2021uw,Goldberg:2025aa,Denis:2026aa}, with concrete applications in polarimetry and magnetometry~\cite{Goldberg:2021tx,Goldberg:2022ac}.

Beyond quantum information, Majorana constellations have been employed to study geometric phases~\cite{Bruno:2012aa,Liu:2014aa}, spinor Bose gases~\cite{Barnett:2006aa,Barnett:2007aa,Makela:2007aa,Lian:2012aa,Cui:2013aa,Serrano-Ensastiga:2023aa}, the Lipkin-Meshkov-Glick model~\cite{Ribeiro:2007aa,Ribeiro:2008aa}, and the symmetries and properties of spin states~\cite{Kolenderski:2008mo,Crann:2010qd,Bannai:2011pi,Baguette:2014ws,Liu:2017aa,Chryssomalakos:2018aa,Wang:2022aa}.

The purpose of this paper is to provide a comprehensive introduction to Majorana constellations, highlighting their broad utility and explanatory power.

\section{Majorana constellations}

We consider  pure spin-$S$ states living in the $(2S+1)$-dimensional Hilbert space $\mathcal{H}_{S}$, carrier of the irreducible representation (irrep) of spin $S$ of SU(2). This space $\mathcal{H}_{S}$ is spanned by the standard angular momentum basis $\{ |S, m\rangle \mid m= -S, \ldots, S\}$, whose elements are simultaneous eigenstates of ${\mathbf{S}}^{2}$ and $ {S}_{z}$. Although $\mathcal{H}_{S}$ is isomorphic to $\mathbb{C}^{2S+1}$, physical states correspond to rays in this space (i.e., vectors differing by only a global phase represent the same state), so the set of pure spin-$S$ states forms the complex projective space $\mathbb{C}\mathbf{P}^{2S}$~\cite{Bengtsson:2017aa}. 

The merit of Majorana was to show that points in $\mathbb{C}\mathbf{P}^{2S}$ are in one-to-one correspondence with unordered sets of (possibly coincident) $2S$ points on the unit sphere $\mathcal{S}_{2}$. In other words,  spin-$S$ states can be obtained as fully symmetrized states of a system of $2S$ spins $1/2$ (or qubits). There are various ways to see why this is so~\cite{Bacry:2004aa}, but probably the most direct  one  is in terms of coherent states.

The spin (or Bloch) coherent states live in $\mathcal{H}_{S}$ and are displaced versions of a fiducial state, much the same as for the canonical coherent states on the plane. This fiducial state is chosen so as to minimize the variance of the Casimir operator $\mathbf{S}^{2} = S_{x}^{2} + S_{y}^{2} + S_{z}^{2}$, where $(S_{x}, S_{y}, S_{z})$ are the angular momentum operators, which generate the algebra $\mathfrak{su}(2)$. The minimum-variance states are $|S, \pm S \rangle$ and they guarantee that their displaced versions are the closest to classical states. The displacement operator on $\mathcal{S}_{2}$ is 
\begin{equation}
D(\theta, \phi) = e^{i \phi S_{z}} e^{i \theta S_{y}} = \exp [ \tfrac{1}{2} \theta (S_{+} e^{-i \phi} - S_{-} e^{i \phi})] \, ,
\end{equation} 
where $S_{\pm} = S_{x} \pm i S_{y}$ are raising and lowering operators. Disentangling this  displacement allows us to express the coherent states $|\theta, \phi \rangle = D(\theta, \phi) |S, -S\rangle$  as~\cite{Perelomov:1986aa,Gazeau:2009aa}
\begin{equation}
|\theta, \phi \rangle \equiv |z \rangle = \frac{1}{( 1 + \lvert z \rvert^{2})^{S}} 
\exp (z {S}_{+}) \ket{S, -S} \, ,
\end{equation}
where the label
\begin{equation}
  z =  \tan \left ( \frac{\theta}{2} \right ) e^{- i \phi}
\end{equation}
corresponds to a stereographic projection from the south pole, mapping the point $(\theta, \phi) \in \mathcal{S}_{2}$ onto the point $z \in \mathbb{C}$~\cite{Coxeter:1969aa}. 

On expanding the exponential, we can write the coherent states in terms of the basis states of the irrep:
\begin{equation}
|z \rangle = \frac{1}{( 1 + \lvert z \rvert^{2})^{S}} \sum_{m=-S}^{S} 
\binom{2S}{S+m}^{\tfrac{1}{2}} \, z^{S+m} \ket{S, m} \, ,
\end{equation} 
or, employing again the stereographic projection,  
\begin{align}
|\theta, \phi \rangle  & =  \sum_{m=-S}^{S} 
\binom{2S}{S+m}^{\tfrac{1}{2}}\; [\sin (\theta/2) ]^{S+m}   
[\cos (\theta/2) ]^{S-m} \nonumber \\
& e^{-i (S+m) \phi} \ket{S, m} \, .
\end{align} 

The system of spin coherent states is complete, but the states are not mutually orthogonal; their overlap is
\begin{equation}
\label{eq:overlap}
 \langle z | z^{\prime} \rangle =
 \frac{(1 + z^{\ast} z^{\prime})^{2S}}{\left [( 1 + \lvert z \rvert^{2}) ( 1 + \lvert z^{\prime} \rvert^{2}) \right ]^{S}} \, .
\end{equation} 
Still, they allow for a resolution of the identity in the form
\begin{equation}
\int_{\mathbb{C}} d\mu_{S} (z) \, \ket{z} \bra{z} = \openone  \, ,
\end{equation}
with the invariant measure given by 
\begin{equation}
d\mu_{S} (z) = \frac{2S+1}{\pi} \frac{d^{2} z}{(1 + \lvert z \rvert^{2})^{2}} \, .
\end{equation} 

The importance of this completeness relation is that it allows one  to decompose an arbitrary pure state over the coherent states.  We denote such a coherent-state representation as $\psi (z^{\ast}) = \langle z | \psi \rangle$. In addition, we define the stellar function $f_\psi (z)$ of the state $|\psi\rangle$ as  
\begin{align}
f_{\psi} (z) & = ( 1 + \lvert z \rvert^{2})^{S} \psi(z) = 
\sum_{m=-S}^{S} \binom{2S}{S+m}^{\tfrac{1}{2}} \psi_{m} \; z^{S+m} \nonumber \\
& = \sum_{k=0}^{2S} \binom{2S}{k}^{\tfrac{1}{2}} \psi_{k-S} \; z^{k}  \, ,
\end{align}
with $\psi_{m} = \langle S,m | \psi\rangle$; in the second line we have made the relabeling $S+m \mapsto k$. Interestingly, in this representation the wave function is a polynomial in $z$ of  order $r \le 2S$. In consequence, the roots $z_{k} \in \mathbb{C}$ of $f_{\psi}$ fully characterize the state:
\begin{equation}
f_\psi ( z ) = \mathcal{N} \prod_{k=1}^{r} (z  - z_{k}) \, ,
\label{eq:fact}
\end{equation}
where $\mathcal{N}$ is an unessential global constant.  These roots define, via an inverse stereographic map, $2S$ points on the unit sphere $\mathcal{S}_{2}$. This is the Majorana constellation, and each one of these points constitutes one star in the constellation.  An SU(2) transformation corresponds to a rigid rotation of the constellation; therefore, states with the same constellation, irrespective of their  orientation, have the same physical properties.

The stellar function is directly related to the Husimi $Q$ function~\cite{Husimi:1940aa,Kano:1965aa}:  
\begin{equation}
Q_\psi ( z ) =   ( 1 + \lvert z \rvert^{2})^{- 2S} \lvert f_{\psi} ( z^{\ast} ) \rvert^{2}  \, ,
\end{equation}
which clearly shows that the zeros of the Husimi $Q_\psi$ function are the complex conjugates of the zeros of $f_{\psi}$.

\section{Extracting information from the  constellation}

So far, we have shown how to compute the constellation when the state is given. In this section, we attack the inverse problem; that is, what information can we extract from a given constellation? To this end, we first recall that every polynomial $P(z)$ of degree $n$ can be represented in terms of its zeros $z_{k}$ using the classical Vieta formulas~\cite{Funkhouser:1930aa}, which can be expressed  in the form
	\begin{equation}
	\label{eq:polzeros1}
		P(z) = \sum_{k=0}^{n} a_{k} z^{k} = a_n \sum_{k=0}^{n} (-1)^{n-k} 
		e_{n-k}(\bm{z} ) \; z^{k} \, ,
	\end{equation}
        where $e_{j} (\bm{z}) \equiv e_{j} (z_{1}, z_{2}, \ldots, z_{n}) $ are the elementary symmetric polynomials~\cite{Waerden:1991aa} defined as
\begin{align}
e_{0} (z_{1}, z_{2}, \ldots, z_{n}) & = 1 \,  , \nonumber \\
e_{1} (z_{1}, z_{2}, \ldots, z_{n}) & = \sum_{1 \le j \le n} z_{j} \,  ,\nonumber \\
e_{2} (z_{1}, z_{2}, \ldots, z_{n}) & = \sum_{1 \le j < k \le n} z_{j} z_{k} \, ,  \nonumber \\
& \vdots \nonumber \\
e_{n} (z_{1}, z_{2}, \ldots, z_{n}) & = z_{1} z_{2} \ldots z_{n} \, .
 \end{align} 
Using this fundamental result, the Majorana stellar function  can be expressed in the compact form   	
\begin{equation}
\label{eq:Majpolbo}
f_\psi (z ) = \sum_{k=0}^{2S}  \mathfrak{f}_{k} (\bm{z}) \; z^{k} \, ,
\end{equation}
where 
\begin{equation}
\label{eq:defpsik}
		\mathfrak{f}_{k} (\bm{z}) =  (-1)^{2S-k} \, \psi_{S} \, e_{2S-k}( \bm{z}) \, ,
	\end{equation}
and the coefficient $\psi_{S}$ is fixed by the normalization condition
	\begin{equation}
\psi_{S}(\bm{z})=\left [ \sum_{k=0}^{2S}\frac{\left|{e_{2S-k}(\bm{z})}\right|^2}{\binom{2S}{k}}\right ]^{- \frac{1}{2}} \, .
	\end{equation}
In this way, the state coefficients are simply related to the coefficients of the Majorana polynomial, and we can calculate them as a function of the stars:
\begin{equation}
\displaystyle
\psi_{k-S}(\bm{z})=\frac{\mathfrak{f}_{k}(\bm{z})}{\binom{2S}{k}^{\frac{1}{2}}} \, .
\end{equation}

\section{Constellations and quantumness}

Let us examine a few relevant examples to illustrate how these constellations look. For a spin coherent state $|z_{0} \rangle$, the stellar representation is direct from Eq.~\eqref{eq:overlap}: 
\begin{equation}
f_{z_{0}} (z)  = \frac{(1 + z_{0}  z)^{2S}}
{(1 + | z_{0} |^{2})^{S}} \, ,
\end{equation}
so it has  a single zero at $z= - 1/z_{0}$ with multiplicity $2S$. In consequence, the constellation collapses in this case to a single point diametrically opposed to the maximum $z_{0}$.

Next, we consider the time-honored NOON states~\cite{Dowling:2008aa}, which, when expressed in the $\ket{S,m}$ basis, take the form   
\begin{equation}
|\psi_{\mathrm{NOON}} \rangle =\frac{1}{\sqrt{2}}(\ket{S,S}+ \ket{S,-S}) \, .
\label{eq:NOON1}
\end{equation}
They are known to have the highest sensitivity for a fixed excitation $S$ to small rotations about the ${z}$-axis~\cite{Bollinger:1996aa}. The associated stellar function reads
\begin{equation}
f_{\mathrm{NOON}} (z) = \frac{1}{\sqrt{2}}  (z^{2S} - 1)\, .
\end{equation}
The zeros are thus the $2S$ roots of unity, so the Majorana constellations have $2S$ stars placed around the equator  with equal angular separation between each star (that is, a regular $2S$gon on the equator). A rotation around the $z$ axis of angle $\pi/(2 S)$ renders the state orthogonal to itself, justifying their optimality.

These two examples suggest that the most classical states correspond to the most concentrated constellations. As the stars spread out and form more symmetric patterns, the state acquires a more pronounced quantum character. In other words, the geometry and symmetry of the constellation directly encode the quantumness of the state. 

To quantify this connection, it is convenient to use the concept of state multipoles, which are the coefficients $\varrho_{Kq}$ in the expansion of the density matrix $\varrho$ in terms of irreducible tensors~\cite{Fano:1959ly,Blum:1981ya} $T_{Kq}$ (where $K=0, \ldots, 2S$ and $q= -K, \ldots, K$).  These tensors form an orthonormal basis and have the correct transformation properties under SU(2) actions.  However, we do not need the whole machinery of irreducible tensor analysis to grasp their physical meaning. Indeed, one can show that~\cite{Goldberg:2021tx} 
\begin{equation}
  \varrho_{Kq}  = \mathcal{C}_{K} 
   \int_{\mathcal{S}_2}  d\Omega \;  Q( \theta, \phi) \;  {Y}_{Kq} (\theta, \phi) \, ,
 \end{equation}
 where $\mathcal{C}_{K}$ is a constant, $d\Omega = \sin \theta\ d\theta\ d\phi $ is the invariant measure on $\mathcal{S}_{2}$ and $Y_{Kq}(\theta, \phi)$ are the spherical harmonics~\cite{Varshalovich:1988ct},  which form an orthonormal basis for square-integrable functions defined on the sphere. The key point is that  the state multipoles appear as the familiar multipoles  in electrostatics~\cite{Jackson:1999aa}, except that the charge density is replaced by $Q( \theta, \phi)$  and distances are replaced by directions.

When expressed in a  Cartesian basis these multipoles appear in a very transparent form. For instance, the three dipole $(\varrho_{1q})$ and the five quadrupole $(\varrho_{2q})$ terms can be written, respectively, as
\begin{equation}
 \mathfrak{p}_{i}  =  \langle n_{i} \rangle \, , \qquad  \qquad 
\mathfrak{q}_{ij}=\langle 3 n_{i} n_{j} - 
\delta_{ij} \rangle \, ,
 \end{equation} 
where  $i, j \in \{x, y, z\}$, $\mathbf{n}$ is a unit vector in $\mathcal{S}_{2}$, and the expectation values of a function $f (\mathbf{n})$ are calculated with respect the $Q$-function
\begin{equation}
 \langle f ( \theta, \phi ) \rangle=  \frac{\int_{\mathcal{S}_2}  d \Omega  \;
f( \theta, \phi ) \; Q(  \theta, \phi)}
{\int_{\mathcal{S}_2} d\Omega  \;  Q( \theta, \phi)} \, .
\end{equation}
This confirms that the multipoles correspond to the $K$th directional moments of the state, resolving progressively finer angular features. The dipole term, as the first-order moment, reproduces the classical picture, while higher-order multipoles describe finer quantum fluctuations. The vanishing of a given $K$th multipole signals that the associated fluctuations are isotropic.

We note that the length of  $\varrho_{Kq} $, defined as
\begin{equation}
\mathfrak{w}_{K} = \sum_{q=- K}^{K}  \lvert \varrho_{Kq} \rvert ^{2}  \, ,
\end{equation}
is just the state overlapping with the $K$th multipole pattern.  For most states only a limited number of multipoles play a substantive role and the rest of them have an exceedingly small contribution.  Therefore, it seems that a convenient way to embody the information is to look at the cumulative distribution
\begin{equation}
  \label{eq:cum}
  \mathcal{A}_{M} = \sum_{K= 1}^{M} \mathfrak{w}_{K} = \sum_{K= 1}^{M}  \sum_{q=- K}^{K}  \lvert \varrho_{Kq} \rvert^{2}  \, ,
\end{equation}
which conveys the whole  information \textit{up} to order  $M$.  The monopole term $|\varrho_{00}|^2$ is excluded, for it is just a constant term.  As with any cumulative distribution, $ \mathcal{A}_{M} $ is a monotone nondecreasing function of the multipole order.

\begin{figure}[t]
\centering 
\includegraphics[width=\columnwidth]{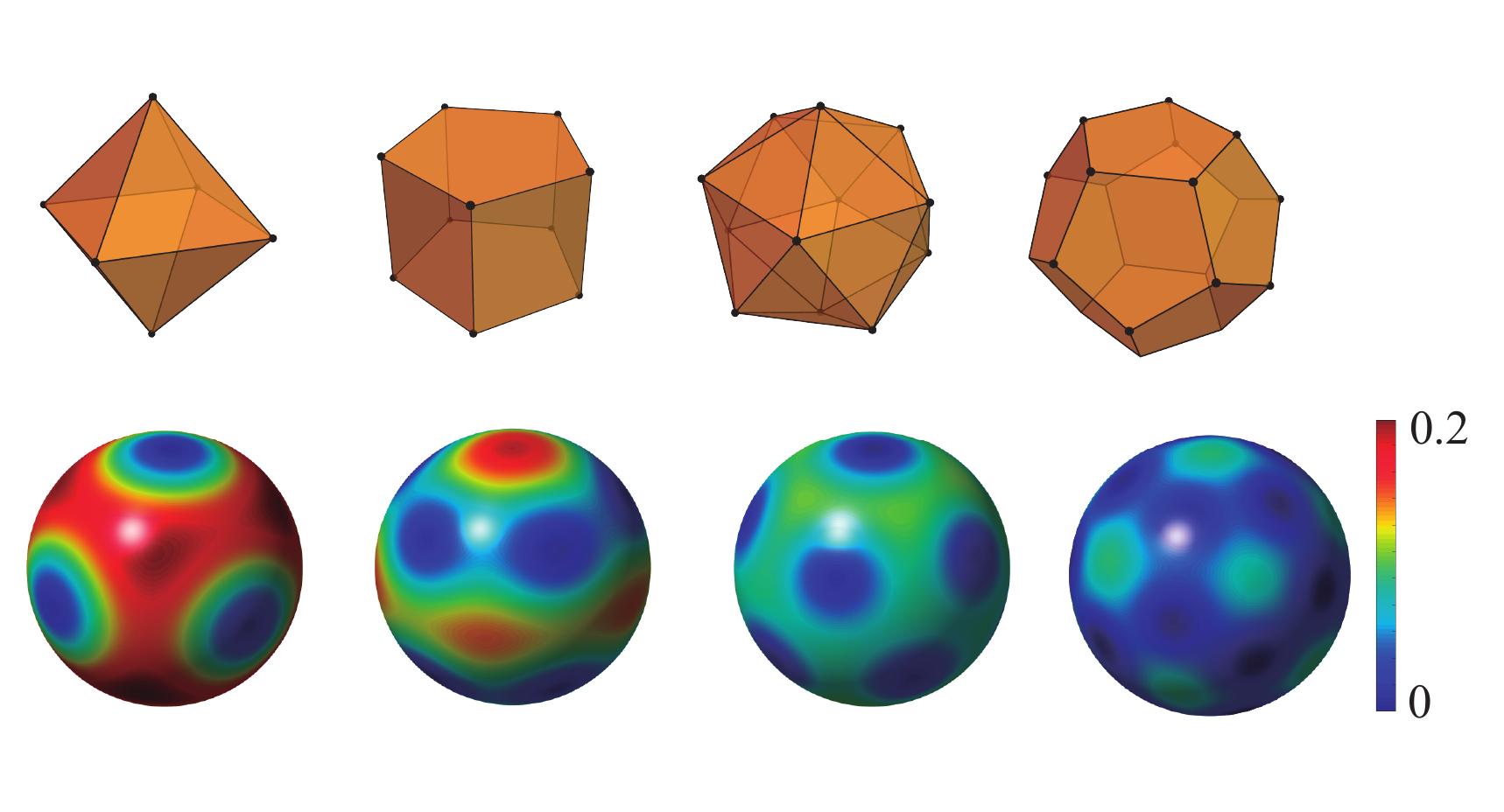}
\caption{Density plots of the  $Q$ functions for the optimal Kings of Quantumness for the cases  
$S=3, 5, 6,$ and 10 (from left to right, blue indicates the zero values and red maximal 	ones). On top, we sketch the Majorana constellations.}  
\label{fig:const}
\end{figure} 

This aggregated multipole strength  $ \mathcal{A}_{M}$ is a good measure of quantumness. It has been proven~\cite{Bjork:2015ux} that is maximal for coherent states. It is thus irresistible to ask which states attain the minimum of this magnitude, as they can be considered in a sense as ``the opposite''  of SU(2) coherent states and so the most quantum ones. For a given $S$ one cannot find pure states that have isotropic moments up to order $M=2S$.  Only the completely mixed states have this property~\cite{Prakash:1971aa,Agarwal:1971aa,Soderholm:2001aa}.   Thus, for each $S$ there exists a set of pure states that are maximally unpolarized and the task is to find what the maximal degree $M$ of ``unpolarization'' can be, and what these states are.    The resulting states are called the Kings of Quantumness and have been fully described in Ref.~\cite{Grassl:2015aa} and experimentally generated~\cite{Bouchard:2017aa,Goldberg:2025aa}. A few associated constellations appear in Fig.~\ref{fig:const}.

 In the framework of the Majorana representation, the search for the most quantum states can be reformulated as the problem of distributing $N$ points on  a sphere in the ``most symmetric''  possible way. This problem has a long history and has inspired a variety of approaches, depending on the particular cost function one aims to optimize~\cite{Conway:1996ys,Saff:1997aa}. This is closely related to relevant problems, such as spherical $t$-designs~\cite{Delsarte:1977dn,Hardin:1992mi,Hardin:1996bv},   the Thomson problem~\cite{Thomson:1904qp,Ashby:1986bk,Edmundson:1992uf,Melnyk:1977gm}, or the notion of  $k$-maximally mixed states~\cite{Arnaud:2013hm}.
  
\section{Stellar dynamics}

It is well established that the group SU(2) acts on the vector space $\mathcal{F}_{S} (\mathbb{C})$ of complex polynomials of degree at most $2S+1$  via the differential realization~\cite{Klimov:2009aa}
\begin{equation}
S_{+} \mapsto  2S z - z^{2} \partial_{z} \, , \qquad
S_{-} \mapsto \partial_{z}  \, , \qquad
S_{z}  \mapsto  z \partial_{z}  - S \, ,
\end{equation}
where $\partial_{z}$ denotes derivative with respect to $z$.   Accordingly, the action of an operator $A$ on this space can be expressed as 
\begin{equation}
A \mapsto A(z, \partial_{z} ) = \sum_{n} a_{n} (z )  \; \partial_{z}^{n} \, ,
\label{eq:map}
\end{equation}
with $a_{n} (z )$ being polynomial functions that depend on the  operator $A$. This representation is valid  for a broad class of operators and, in particular, for all the elements in the enveloping algebra of $\mathfrak{su}(2)$.  The highest meaningful order $n$ in this equation is $2S$ since $\partial_{z^{\ast}} ^n \psi (z^{\ast}) = 0$ for $n > 2S$.

Since, as we have discussed before, the coherent-state wave function $\psi(z^{\ast})$ belongs to $\mathcal{F}_{S} (\mathbb{C})$, the Schr\"{o}dinger equation can be recast as (setting $\hbar=1$)
\begin{equation}
i \partial_{t}{\psi} (z^{\ast}, t) = H(z^{\ast}, \partial_{z^{\ast}}) \psi ( z^{\ast}, t) \, . 
\label{eq:Schr}
\end{equation}   
If at time $t$, the state has a zero at some complex point $z_{k}^{\ast}$, then the condition  that the zero persists under an infinitesimal time evolution, $\psi (z_{k}^{\ast} + \delta z_{k}^{\ast}, t +\delta t) = 0$, implies
\begin{equation}
	\dot{z}_{k}^{\ast}  =  \frac{\delta z^{\ast}_{k}}{\delta t} =  -  \left . \frac{\partial_t  \psi}{\partial_{z^{\ast}} \psi} \right |_{z^{\ast}_{k}} \, ,
	\label{eq:evol1}
\end{equation} 
which, after using \eqref{eq:Schr}, becomes
\begin{equation}
  \dot{z}_{k}^{\ast} = i  \left . \frac{H(z^{\ast}, \partial_{z^{\ast}}) \psi (z^{\ast})}{\partial_{z^{\ast}} \psi (z^{\ast})} \right |_{z^{\ast}_{k}} \, ,
  \label{eq:evol2}
\end{equation} 
These equations determine the dynamics of each of the Majorana stars associated with a given state~\cite{Leboeuf:1990aa,Leboeuf:1991aa}.  However, they are not  yet closed, as the wave function  $\psi ( z^\ast , t)$ still appears explicitly  when calculating  $\dot{z}_{k}^{\ast}$. To eliminate this dependence, we write  
\begin{equation}
  \left . \frac{H(z^{\ast}, \partial_{z^{\ast}}) \psi (z^{\ast})}{\partial_{z^{\ast}} \psi (z^{\ast})} \right |_{z_{k}^{\ast}} = \left . \sum_{n}  h_{n} (z^{\ast}_{k}) \frac{\partial_{z^{\ast}}^{n} \psi}{\partial_{z^{\ast}} \psi} \right |_{z_{k}^{\ast}}   \, ,
	\label{eq:evol3}
\end{equation} 
where $h_n$ are the polynomial coefficients associated with the Hamiltonian  [see Eq.~\eqref{eq:map}]. Using the factorization property~\eqref{eq:fact}, the ratio of derivatives can be expressed as
\begin{equation}
\left . 	\frac{\partial_{z^{\ast}}^{n} \psi}{\partial_{z^{\ast}} \psi}  \right |_{z_{k}^{\ast}}  = \sum_{k_1 < k_2 < \ldots < k_{n-1}}^{2S}  
	\frac{n!}{(z_{k}^{\ast} -z_{k_1}^{\ast}) \ldots (z_{k}^{\ast}-z_{k_{n-1}}^{\ast})}\, ,
\label{eq:evol4}
\end{equation}
where the sum runs  over terms with all $k_i \neq k$. The final form for the equations of motion thus reads
\begin{equation} 
	  \dot{z}_{k}^{\ast} =	i \sum_{n} h_{n}  (z^{\ast}_{k})  \sum_{k_1 < k_2 < \ldots < k_{n-1}}^{2S}  
	\frac{n!}{(z_{k}^{\ast} -z_{k_1}^{\ast}) \ldots (z_{k}^{\ast}-z_{k_{n-1}}^{\ast})}\, ,
	\label{eq:evolfin}
\end{equation}
The equilibrium configurations for the zeros are determined by the conditions $\dot{z}_{k}^{\ast}  = 0$. This gives an algebraic system of   $2S$ nonlinear coupled equations for the $2S$ variables  $\{ z_k^\ast\}$. 

Each term in the sum corresponds to an effective $n$-body interaction: the $k$th star interacts with $n-1$ of the remaining  stars. The number of such contributions is $C(2S-2,n-2)$, where $C(n,k)$ denotes a binomial coefficient.  Then the coefficient $h_n $ can be seen as a position-dependent \emph{charge}. However, this charge depends only on the position of the $k$th zero, and not on the position of the remaining zeros. 

For degenerate stars, both $\partial_{z^{\ast}} \psi$ and  $\partial_{t} \psi$ vanish. The evolution equations for the stars are then found from higher-order expansions in $\delta z^{\ast}$ and $\delta t$, yielding  coupled equations for $m$th-order degenerate stars 
\begin{equation}
\left . 	\sum_{\ell=0}^{m} z^{\ast \ell}  {{m}\choose{\ell}} \partial_{z^{\ast}}^{\ell}   [- i H(z^{\ast}, \partial_{z^{\ast}} )^{m-\ell} ] \psi \right |_{z_{k}^{\ast}}  = 0 \, .
\end{equation} 

We now illustrate our approach with a few simple examples. We begin with the linear Hamiltonian $H = \omega_{0} S_{z}$. In the coherent-state representation, the symbol of this Hamiltonian reads
\begin{equation}
H = \omega_{0} (z^{\ast}  \partial_{z^{\ast}} - S) \, .
\end{equation}
Applying now \eqref{eq:evol1}, the dynamics of the stars is governed by
\begin{equation}
\dot{z}^{\ast}_{k} = i \omega_{0} \, z^{\ast}_{k} \, ,
\end{equation}
which leads to the straightforward solution
\begin{equation}
z^{\ast}_{k} (t) = z^{\ast}_{k} e^{i \omega_{0} t } \, .
\end{equation}
This means that the all the stars rotate around the $z$ axis with the same velocity, exactly as a classical state. 

Next, we consider the Kerr Hamiltonian $H= \chi S_{z}^{2}$~\cite{Klimov:2009aa}, with analytic realization
\begin{equation}
H = \chi [z^{\ast \, 2}  \partial_{z^{\ast}}^{ 2} -  (2S-1) z^{\ast}  \partial_{z^{\ast}} + S^{2} ] \, .
\end{equation}
Again \eqref{eq:evol1} immediately yields 
\begin{equation}
\dot{z}^{\ast}_{k} = i \chi \left [ z_{k}^{\ast \, 2} \sum_{k\neq \ell}^{2S} \frac{1}{z^{\ast}_{k} - z^{\ast}_{\ell}} - \left ( S - \frac{1}{2} \right ) \right ] \, .
\end{equation}
This nonlinear Hamiltonian induces interactions between the stars: points located at different latitudes move with different velocities and/or in different directions, producing a deformation of the constellation~\cite{Leboeuf:1991aa}. 

\bigskip

\section{Concluding remarks}

The Majorana stellar representation stands out as an elegant and insightful way to visualize quantum states beyond the spin-$1/2$ case. By expressing a spin-$S$ state as a constellation of $2S$ points on the Bloch sphere, it transforms the abstract structure of high-dimensional Hilbert spaces into an intuitive geometric picture. This geometric perspective deepens our understanding of quantum superposition, coherence, and symmetry, providing a natural language to discuss phenomena such as spin squeezing, entanglement, and state evolution. Moreover, the motion and arrangement of the Majorana stars reveal subtle features of quantum dynamics and interference that are often hidden in algebraic formulations. 

Owing to its unifying and visual character, the Majorana representation continues to inspire new insights across quantum optics, quantum information, and the broader study of many-body and topological quantum systems.

\section*{ackowledgments}
The ideas presented in this paper have been shaped and enriched through the questions, suggestions, criticism, and advice of numerous colleagues. We extend our special gratitude to G. Björk, P. de la Hoz,  M. Grassl, H. de Guise, A. Mu\~{n}oz, and J. L. R. Herv\'as  for their valuable contributions in various ways. This work is supported by the Spanish Agencia Estatal de Investigaci\'on  (Grant PID2021-127781NB-I00) and the Mexican Consejo Nacional de Humanidades, Ciencias y Tecnologías (Grant CBF2023-2024-50).

%

\end{document}